\documentclass[12pt,aps,prb,preprint]{revtex4}   

\usepackage{amsmath}    
\usepackage{graphicx}   
\usepackage{natbib}

\begin{document}

\title{Emphasizing Expert Practice with Spaced Recall}
\author{Eugene T. Torigoe}
 \affiliation{Thiel College, Greenville, PA 16125}
 \email{etorigoe@thiel.edu}   
\date{\today}

\begin{abstract}
This paper explores an intervention that emphasizes expert practice using spaced recall.  Interviews were performed with two students who were shown physics solutions, and were asked to repeatedly recall the solutions over a period of weeks.  The students reflected that they became aware of the importance of using the diagrams to create equations, as well as the utility of reasoning over pure memorization.  We believe that the structure of this activity may be an effective way of encouraging expert practice to introductory physics students.
\end{abstract}

\maketitle

\section{Introduction}
Spaced recall has been validated over many decades of psychological research as an effective technique to promote learning and retention.~\cite{Brown-14, Roediger-06}  The traditional assumption in education is that students learn from lectures, readings, and studying, which can then be measured by later tests.  But this research has shown that the recall of information, as on a test, is a highly effective method of learning and retaining information long-term.  In fact, spaced recall has been shown to be more effective a technique for long-term retention than repeated study.  "The present research shows the powerful effect of testing on learning: Repeated retrieval practice enhanced long-term retention, whereas repeated studying produced essentially no benefit."~\cite{Karpicke-08}

The effectiveness of spaced recall depends on the difficulty and effort expended by the student to recall the information.  The psychological researchers refer to this as "desirable difficulty", because of the benefit of struggle.  Even though students may very successfully recall material immediately after presentation or studying, this type of recall is not effective because it relies on the short-term memory.  Spaced recall relies on the time between the recall sessions (days or weeks), which increases the cognitive difficulty for the student.  The more spaced recall sessions occur the greater the benefit.

In this paper we propose a methodology to incorporate spaced recall into introductory physics.  We believe that this technique may have multiple benefits.  First, it will introduce students to spaced recall, which is a highly effective method of learning.  In addition, it will help place emphasis on expert practices in physics.  In particular, translating between diagrams and equations, and using symbolic physics equations.

As a research tool, recall sessions are also an interesting way of highlighting what details students use to organize and recall information.  By asking them to recall we can gain insight into what details students pay attention to, which they ignore, and the types of information they have difficulty or ease recalling.  

Prior physics education research has demonstrated that experts organize information differently than novices.  Specifically in physics, experts are more likely to focus on "deep structure", while novices are more likely to focus on "surface features".~\cite{Chi-81, Larkin-80}  Such differences may make the encoding of information for later retrieval difficult in similar circumstances, but especially in novel circumstances.  

Further, there is a great deal of research on the difficulties novices have coordinating multiple representations.~\cite{Kohl-08, Kohl-05, Nieminen-10, VonKorff-12, VanHeuvelen-01, Fredlund-14}  The coordination of representations has been shown to be important for learning physics.  In studies of how students use multiple representations, students in courses that emphasize multiple representations use many representations, but do not use them to the extent that experts use them.~\cite{Kohl-08}  The representations used in physics are often densely packed, and require unpacking for a novice to understand.~\cite{Fredlund-14}  

The diagrams, symbols, and other notations used in introductory physics are culturally specific to the discipline of physics.  There are many examples that the math used in math class is different from the math used in physics classes.~\cite{Redish-15}  When students take introductory physics, they must learn these discipline specific cultural differences to be successful.  Many students in introductory physics also have difficulty using and understanding symbolic physics problems.~\cite{Torigoe-11}

Our hypothesis is that using spaced recall in introductory physics will place a stronger emphasis on practices shared by physics experts, such as symbolic mathematics, and translating between representations.

\section{Methodology}

Two college students who had completed introductory physics the previous year were interviewed 4 times over the course of 3 months.  The students were tasked with reviewing three different problems: Heat Engine, Pendulum, and Wave modes problems (Appendix 1).  During the first interview they watched a video solution, and then were asked to immediately recall the solution they had just seen.  In subsequent meetings they were asked to recall the solutions.  If they became stuck they could then look at their notes from the first meeting.  In the last meeting they were given an alternative version of the problems after recalling the related problem.

The three problems were chosen based on their representational density.  In particular, the solution is aided if the students are able to translate a diagram into the appropriate equation.  The full problems and their solutions can be found in Appendix 1.

\textbf{Heat Engine Problem:} In this problem the students must use the diagram to develop two different equations.  The students must also replace the quantities of the general equation efficiency = benefit/cost.  They also manipulate symbolic equations.

\textbf{Pendulum Problem:} The students create a triangle diagram to find the components of a force.  The also use a free body diagram to apply Newton's 2nd Law.  They specify Newton's 2nd Law with more specific quantities, and manipulate symbolic equations.

\textbf{Wave Modes Problem:} The students use diagrams of standing wave modes to create equations for the wavelength.  The students must then determine the general symbolic rule from the patterns of the earlier equations.

Alternate versions of each question were made and can be seen in Appendix 2.  The alternative versions of the questions were given to the students during the last interview after they recalled the original versions.

\section{What We Learned from their Mistakes}

Although both students were able to flawlessly recall the solutions to all three problems immediately after watching the solution videos, consistent success at recalling the problems did not occur until after at least one unsuccessful spaced recall session.  In some cases the students made mistakes on problems they had solved perfectly the previous week.  Although usually these mistakes were minor errors.  So while there was a definite trend of improvement over time, it was not monotonic.  

It was clear from where the students got stuck and from their comments that one of the most memorable features of the solution was that the equations were to be combined by substitution.  In the 2nd interview during the first long term recall session both had difficulty recalling the heat engine solution.  One of the students made the following remarks of what they could remember.

\begin{quote}\emph{
I'm kind of stuck.  So I know that there's two equations, and the first one you use for the 2nd equation to simplify it.  But I can't remember what they are.}
\end{quote}

Even though neither were able to recall the entire solution for any of the problems during the first spaced recall interview, they both generally did a good job of remembering the diagram and how to correctly label it.  They both felt that remembering diagrams was much easier than remembering the equations.  

\begin{quote}\emph{
I was pretty sure that I would remember the picture, because I remember pictures easier, but the equations are usually what I struggle on.}
\end{quote}

One particular exception to the accuracy of the pictures was in the heat engine diagram.  Neither student in any of the recall sessions (including the immediate recall) used the width of the bars to represent the relative magnitude of the heat flows and work, even though the solution video presents that information.  This feature of the diagram was obviously not a very salient feature.  One of the students replaced these bars with arrows during the first spaced recall and in subsequent interviews (Figure 1).

\begin{figure}
\begin{center}
\scalebox{0.5}{\includegraphics{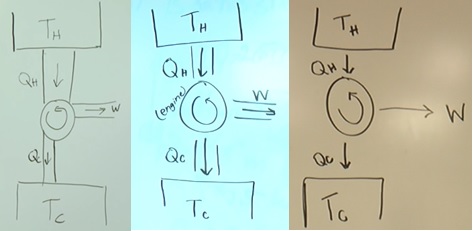}}
\caption{\label{fig:HeatEngines} (Left) The diagram from the solution video, (Middle) The diagram from one of the students that does not represent relative magnitude using the width of the bars, (Right) The diagram from the other student that replaces the bars with arrows.}
\end{center}
\end{figure}

It is interesting to note that during the immediate recall after viewing the correct solution, which both recalled correctly for all of the problems, they referenced the diagrams when asked to explain the source of the equations they used.  But, when attempting to recall the solution during the first spaced recall session 2 weeks later neither attempted to use the diagram to derive the equation.  This could perhaps be interpreted to mean that the students saw the diagrams as a method to communicate the solution, but not to actually generate the equations during problem solving.

After their first unsuccessful recall session, the interviewer encouraged both students to focus more on using the diagram to derive the correct equation.  In the last interview both referenced the importance of this advice in their success in the following weeks.

\section{Learning to translate between the pictures and equations}

Both of the subjects remarked during the final interview that they had become much better at being able to translate between the diagrams and the equations.  Our hypothesis for this improvement was a result of the spaced recall structure of the activity.  Even though learning to translate between a diagram and an equation is difficult, it is far easier than memorizing the entire problem.  This activity shifts the burden, and makes translating between representations the smaller encumbrance.

\begin{quotation}\emph{
So tell me, do you think there is any change in the way you use diagrams since when you started this process?
}
\end{quotation}

\begin{quotation}\emph{
Yeah a little more I learned to look into the actual diagram, and look at what it is telling me, not that it's just there as a picture to help explain the problem
}
\end{quotation}

The student had treated the diagram as a representation of the problem to be solved instead of as a tool that could aid in the problem solving process.

\begin{quotation}\emph{
You know I'll tell you that one of the things I noticed, when we did the first two interviews I felt like you weren't kind of using the pictures as much as you could have, in these last two interviews, you memorized the pictures, which is fine, that's actually the way I approach, I memorize the picture, but then I use it to kind of reason through the problem 
}
\end{quotation}
\begin{quotation}\emph{
Yeah I remember how it went through.  I never really looked through diagrams as a tool, I just looked at them as they were just there, that's what it looks like, I never plugged things in and solved an equation with it ... I thought it was interesting, as being the subject, I guess you could say, somehow it taught me how to actually use the diagram, that's one of the main things that I'll pull away from this, if I have a class in the future that uses diagrams
}
\end{quotation}

The other subject remarked how their attitude changed from total memorization to using the diagram to work through the problems.

\begin{quotation}
	\emph{Do you think that there have been any changes in the way you solved these problems, or attempted these problems?
	}
\end{quotation}
\begin{quotation}
	\emph{
I think in the first ones, the original ones [as opposed to the alternate versions], I used the pictures a lot more.  The very first time I tried them I just tried to remember what the answer was ... Not quite the answer, but like all of the work for it, but once I looked at the picture and remembered how to draw the picture I could apply it a lot better.  
}
\end{quotation}

\section{Reasoning vs. Memorization}

We also found evidence that this structure encouraged strategic memorization of certain parts of the solution, and then reasoning through the rest of the solution.  Again even though this is difficult, it is much less difficult than memorizing the entire solution.

\begin{quotation}
	\emph{When you go through these problems, do you feel like you memorize them, or do you feel like you are working your way through them?
}
\end{quotation}
\begin{quotation}
	\emph{
Slightly.  I memorized some of it because, um, I knew that there was, I don't know, I think I memorized it, but when going through it I don't remember how it goes until I look at it a separate way, or I think of it in a different way, and then it reminds me that oh this is how I need to do it
}
\end{quotation}

When we questioned one of the students about the amount of the solution that they memorized, they confirmed that they only memorized the diagram, and then used the diagram to work through the rest of the problem.

\begin{quotation}
	\emph{You basically memorized that heat engine problem?
}
\end{quotation}
\begin{quotation}
	\emph{
I memorized the diagram, and I memorized the benefit over cost, and since I memorized those things it was easier for me to incorporate
}
\end{quotation}
\begin{quotation}
	\emph{
Yeah, but did you feel like you were reasoning your way through them?
}
\end{quotation}
\begin{quotation}
	\emph{
Slightly when it came to how to get the equation to plug into benefit over cost, but other than that I pretty much knew the diagram without having to read the problem
}
\end{quotation}

The same subject made a similar response on the pendulum problem

\begin{quotation}
	\emph{Yeah, so when you are going through this problem, what parts do you think you memorized, and what parts do you think you are reasoning, what parts do you have to think about?
}
\end{quotation}
\begin{quotation}
	\emph{
I pretty much had this memorized [triangle] except switching them [referring to sine and cosine error] I didn't remember what to solve for, but throughout this part [Newton's 2nd Law] I used the reasoning through the diagram 
}
\end{quotation}

When we asked the other subject about the Wave Mode problem, the other subject also said that they reasoned their way through the solution.

\begin{quotation}
	\emph{So, when you are solving these problems, let's say you are solving this kind of problem how much do you think you memorized vs how much are you working your way through?  Obviously you've never seen this problem before [referring to the alternate problem] but 
}
\end{quotation}
\begin{quotation}
	\emph{
Uh I don't, I memorized like half of a node equals a wavelength, but after that I think it's all working through it.  [The student uses the term "node" to refer to a loop in the diagram]
}
\end{quotation}

It was clear from their work as well as their remarks that neither memorized the entire solution.  Both were able to strategically memorize portions of the solution, but then used the diagrams work through the solutions.

\section{Transferring to New Situations}

During the final interview the students were given alternative versions of the problems.  The purpose of these variations was to see if the students would be able to translate a given diagram into an appropriate equation in slightly unfamiliar circumstances.  The students did surprisingly well on these alternate problems.  The students were able to correctly translate the new diagrams, which were given, into the appropriate diagrams in 5 of the 6 cases.  This seems to confirm that the students were able to consistently translate the diagrams into equations.  But even though they could successfully turn information in the diagram into an equation, most of their solutions were either incomplete or contained a minor error of some kind.  

In the alternate version of the pendulum problem, both students were able to correctly draw the free body diagram, and apply Newton's 2nd Law, but both combined the equations even though it was not necessary to solve for the acceleration.  This is additional evidence that combining the equations seemed to be a very memorable feature of the problem's solution.

Both students also were unable to complete the alternate version of the Wave Mode problem because neither could figure out how to use the integer parameter n to create an equation for odd numbers.

\section{Discussion}

Spaced recall of representationally rich physics problem in two interviews showed evidence that such a scheme encourages expert practices in physics.  Memorizing every detail of the solution is very difficult, in part because of the number of steps involved.  In contrast, the subjects developed expert-like habits such as strategic memorization, translating the diagram into an equation, and reasoning through the problem.  

One of the most salient features of the solution for both students was that one equation was to be substituted in for another equation.  This feature seemed even more prominent in their memory than how to derive the equations that were to be combined.

Success recalling a solution occurred immediately after seeing the video solution, and not again until at least the 2nd spaced recall session.  In a typical class, students see a series of examples, which they are expected to have understood and internalized.  But this research makes it seem doubtful that many students could retain the information from a lecture without additional out of class study.  In order for the students to retain the information from the lectures, then they would have to review each lecture multiple times outside of class over a period of days or weeks.  From my experience as a physics instructor, such additional out-of-class work is unlikely to be occurring.

Although this spaced recall methodology was only performed on two students, who had previously passed introductory physics, we believe that it shows promise as an in-class activity for introductory physics students.  

In fall 2015 we will be using spaced recall for a class of 20 students in calculus-based introductory physics.

\appendix

\section{Appendix 1: The Problems and Solutions}

\textbf{Question 1:}
A heat engine operates between a hot thermal reservoir at a temperature TH and a cold thermal reservoir at TC.  Create an energy flow diagram for the thermal reservoirs and the heat engine, and use it to derive a symbolic expression for the thermal efficiency, η, of the heat engine in terms of the heat from the high temperature reservoir QH and the heat from the cold temperature reservoir QC.
 
\begin{figure}
\begin{center}
\scalebox{0.5}{\includegraphics{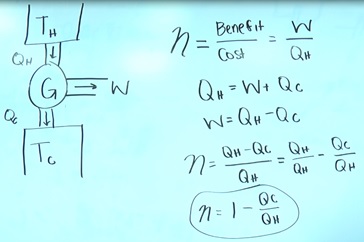}}
\caption{\label{fig:Q1} Solution to the Heat Engine problem..}
\end{center}
\end{figure}

\textbf{Question 2:}
A string of length L is fixed at both ends.  Draw the first 4 standing wave modes, and use the pictures to derive an expression for the wavelength (λ) of the nth standing wave mode (n = 1, 2, 3, 4, …) as a function of the length L.

\begin{figure}
\begin{center}
\scalebox{0.5}{\includegraphics{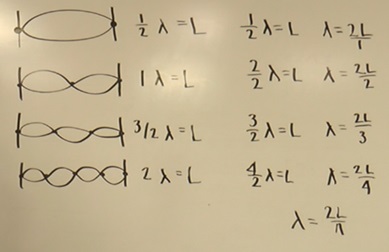}}
\caption{\label{fig:Q2} Solution to the Wave Mode problem.}
\end{center}
\end{figure}

\textbf{Question 3:}
A pendulum (a ball connected to a thin string) has a mass m, and a length L is hung from the rear view mirror of a car.  When the car is at rest the pendulum hangs down in the vertical direction.  When the car has constant acceleration in the forward direction the pendulum is at a stable angle θ with respect to the vertical.  Draw a free body diagram for the mass m, and use Newton’s 2nd Law to find an expression for the acceleration (a) in terms of the angle θ. 

\begin{figure}
\begin{center}
\scalebox{0.5}{\includegraphics{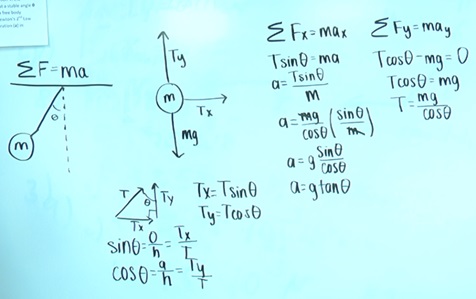}}
\caption{\label{fig:Q3} Solution to the Pendulum problem.}
\end{center}
\end{figure}

\section{Appendix 2: Transfer Problems and Solutions}

\textbf{Question 1 ALT:}
A heat engine operates between a hot thermal reservoir at a temperature T¬H and a cold thermal reservoir at TC.  A furnace (high temp reservoir) sends heat to a heat engine (QH1), as well as directly to the environment (QH2).  The energy flow diagram is shown at right.  Find the expression for the thermal efficiency, η, of the overall factory including the heat engine and the furnace in terms of the heat flows given QC, QH1, and QH2. 
 
\begin{figure}
\begin{center}
\scalebox{0.5}{\includegraphics{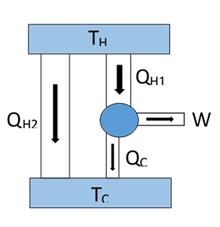}}
\caption{\label{fig:AltQ1} Picture given to the students for Question 1 ALT}
\end{center}
\end{figure}

\begin{figure}
\begin{center}
\scalebox{0.5}{\includegraphics{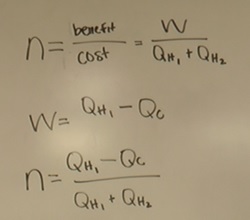}}
\caption{\label{fig:AltQ1sol} Solution to the Alternative Heat Engine problem.}
\end{center}
\end{figure}

\textbf{Question 2 Alt:}
A pipe has a length L is closed on one end and open on the other end.  At right are the first 4 standing wave modes for displacement.   Use the pictures to derive an expression for the wavelength (λ ) of the nth standing wave mode (n = 1, 2, 3, 4, …) as a function of the length L.

\begin{figure}
\begin{center}
\scalebox{0.5}{\includegraphics{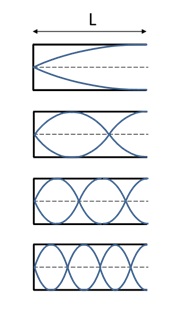}}
\caption{\label{fig:AltQ2} Picture given to the students for Question 1 ALT}
\end{center}
\end{figure}

\begin{figure}
\begin{center}
\scalebox{0.5}{\includegraphics{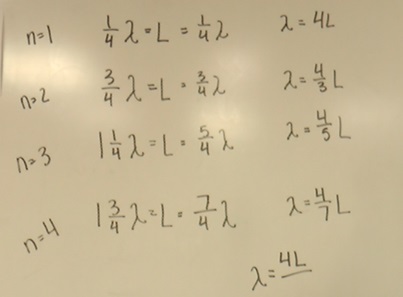}}
\caption{\label{fig:AltQ2sol} Solution to the Alternative Wave Mode problem.  The final expression should be: λ=  4L/(2n-1).  Neither student could figure out how to represent odd numbers using the integer parameter n.}
\end{center}
\end{figure}

\textbf{Question 3 Alt:}
A pendulum (a ball connected to a thin string) has a mass m, and a length L is hung from the rear view mirror of a car.  When the car has constant acceleration down an incline the pendulum is at a stable angle θ with respect to the vertical.  The free body diagram is shown at the bottom right with rotated x and y axes that are parallel and perpendicular to the incline, respectively.  Use Newton’s 2nd Law to find an expression for the acceleration (a) in terms of the angle θ.
 
\begin{figure}
\begin{center}
\scalebox{0.5}{\includegraphics{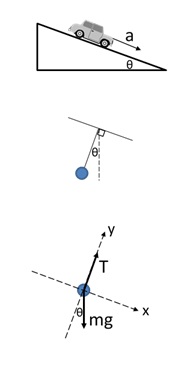}}
\caption{\label{fig:AltQ3} Picture given to the students for Question 1 ALT}
\end{center}
\end{figure}

\begin{figure}
\begin{center}
\scalebox{0.5}{\includegraphics{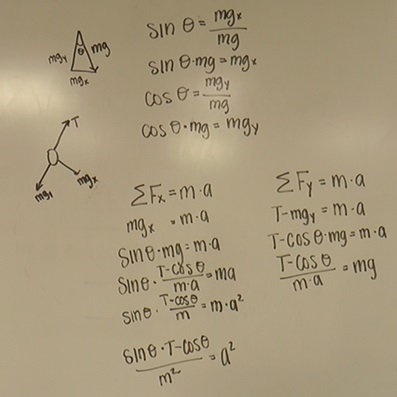}}
\caption{\label{fig:AltQ3sol} Solution to the Alternative Pendulum problem.  The final result should be a = g sinθ.  Both students combined the two equations stemming from the application of Newton’s 2nd Law, even though it was unnecessary.}
\end{center}
\end{figure}

\end{document}